\documentstyle[aps,amsfonts,amssymb,twocolumn,epsfig]{revtex} 

\begin{document}
\draft

\title{The controllable $\pi$ - SQUID}

\author{J.J.A.Baselmans, B.J. van Wees, and T.M. Klapwijk
$^\ast$}
\address{Department of Applied Physics and Materials Science Center, University
of Groningen,\\ Nijenborg 4, 9747 AG Groningen, The Netherlands \\
$^\ast$Department of Applied Physics and DIMES, Delft University
of Technology, \\ Lorentzweg 1, 2628 CJ Delft, The Netherlands \\}
\date{\today}

\maketitle

\begin{abstract}
We have fabricated and studied a new kind of DC SQUID in which the
magnitude and sign of the critical current of the individual
Josephson junctions can be controlled by additional voltage probes
connected to the junctions. We show that the amplitude of the
voltage oscillations of the SQUID as a function of the applied
magnetic field can be tuned and that the phase of the oscillations
can be switched between 0 and $\pi$ in the temperature range of
0.1 - 4.2 K using a suitable control voltage. This is equivalent
to the external application of (n+1/2) flux quantum.

\end{abstract}
\newpage

%

The direct current superconducting quantum interference device (DC
SQUID) is the most sensitive magnetic flux sensor currently
available. It combines two phenomena: Flux(oid) quantization and
the Josephson effect \cite{Clarke,Pavuna,Tinkam}. The critical
current of the SQUID is an oscillatory function of the applied
magnetic flux $\phi_{app}$ with a period given by the flux quantum
$\phi_0 = h/2e = 2.07 *10^{-15}$ Wb. Using a suitable current
$I_{Bias}$ a little larger than the sum of the critical currents
of the 2 Josephson junctions the oscillating critical current is
transformed into an oscillatory voltage. The SQUID can be designed
to meet various demands. However, once fabricated, the properties
of the device, in particular the critical currents of the two
Josephson junctions, are fixed. On the other hand, a recent
development in the field of mesoscopic superconductivity is the
controllable Josepshon junction. In such a junction it is possible
to change the magnitude of the critical current $I_c$
\cite{Alberto,Schapers,Kutchinsky} and even reverse its direction
with respect to the phase difference $\varphi$ between the
superconducting electrodes
\cite{Volkov1,Volkov2,Wilhelm,Yip,me,Delsing}. This corresponds to
an extra phase factor of $\pi$ in the Josephson supercurrent
($I_{sc}$) - phase relation $I_{sc}=I_c sin(\varphi) \Rightarrow
I_{sc}=I_c sin(\varphi+\pi)$. This $\pi$-junction behaviour is
well known in the field of high Tc superconductors
\cite{Harlingen} and has also been observed in ferromagnetic weak
links \cite{Ryazanov}. However, the state (normal or $\pi$) of the
junction is fixed once the device has been made, this in contrast
with a \emph{controllable} $\pi$-junction. We have implemented
such a \emph{controllable} Josephson junction in a DC SQUID, which
leads to a controllable $\pi$-SQUID, in which the critical
currents of the individual junctions \cite{Blamire} and hence the
symmetry of the SQUID can be fine tuned. More interestingly the
device can be switched from a state where no circulating current
is running around the SQUID loop (at $\phi$=n$\phi_0$ with n an
integer), to a state with a circulating current running around the
SQUID loop, without the application of an external magnetic field,
but by switching one of the weak links into the $\pi$-state. This
is a consequence of the condition of a single-valued wave function
around the SQUID loop:
\begin{equation}\label{SQUIDeq}
\frac{2 \pi \phi}{\phi_0}-\varphi_1-\varphi_2= 2 \pi n
\end{equation}
where the total flux $\phi=\phi_{app}+LI$, the flux due to the
screening current. Switching for example junction 1 in the
$\pi$-state changes $\varphi_1$ with $\pi$, which leads to the
same solution of the equation if the junction would be in the
normal state and $\phi=(n+\frac{1}{2})\phi_0$.

In this Letter we propose and demonstrate experimentally the
controllable $\pi$-SQUID and show that the magnitude of the
voltage oscillations as a function of the applied magnetic field
(V-B oscillations) can be tuned and shifted a factor of $\pi$ in
phase.

The only realization so far of a controllable $\pi$-junction is
based upon a superconductor - normal metal - superconductor
junction in which the normal region is made of gold or silver and
the superconductor is made of niobium \cite{me,Delsing}. The
normal region of the junction is connected to the center of a
short mesoscopic wire ($\sim 1 \mu m$), that we will call the
control channel, which is attached also to two large electron
reservoirs. In practice the device has a cross shape, with the
control channel crossing the normal region of the junction. The
principle of operation is the following: A control voltage $V_c$
is applied over the control channel, resulting in a change in the
electron energy distribution in the channel, and therefore the
normal region of the SNS junction. As a consequence the occupation
of the quantum states that carry the supercurrent though the
normal region is also changed. If the control channel is
sufficiently short, so that both electron - phonon and electron -
electron interactions can be neglected, the electron energy
distribution in the center of the control channel will not be a
Fermi-distribution, but the re-normalized superposition of the
electron distribution functions of the two reservoirs. This
distribution is a double step function, with a separation of
$eV_c$ between the steps if $eV_c \gg k_bT$\cite{Pothier}. Here T
is the electron temperature in the reservoirs and $k_b$ is
Boltzmann's constant. The effect of this specific electron
distribution in the normal region of the SNS junction on the
magnitude of the critical current is a reduction of the critical
current to zero and a subsequent sign reversal with increasing
$V_c$. In the limit of strong electron-electron interaction and
still negligible electron-phonon interaction the electron energy
distribution in the control channel will be a thermal one, with
however an elevated effective temperature proportional to $V_c$
(hot electron regime). The effect of such a distribution on the
critical current of the junction is a monotonic decrease to zero
analogous to a rise in temperature \cite{Alberto,menew}.

\begin{figure}
\centerline{\psfig{figure=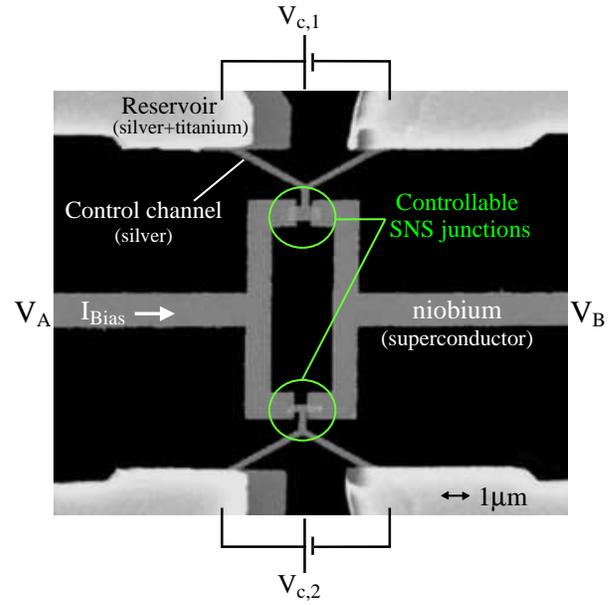,width=8 cm,clip=}}
\caption{SEM-picture of the controllable $\pi$-Squid. The currents
and voltages used in the experiment are also indicated, with the
voltage over the SQUID, $V_{SQUID}$=$V_B$-$V_A$} \label{device}
\end{figure}

A practical realization of a controllable $\pi$-SQUID is shown in
Fig. \ref{device}. A niobium loop (thickness: 50 nm, surface area:
12 $\mu m^2$) has two metallic weak links made of silver
(thickness 50 nm). The length of the normal regions of both
junctions is 1100 nm with a Nb separation of 420 nm. The width of
the normal regions is 520 nm for the top junction and 220 nm for
the bottom junction. The silver weak links are each connected to a
V-shaped silver control channel with a total length of 5+1 $\mu m$
which connects to two large silver reservoirs of 475 nm thick and
a surface area of about $1 mm^2$. The size of the reservoirs is
needed because they should act as effective cooling fins to
prevent unwanted electron heating at
T$<$1K\cite{UrbinaClarke,Henny}. The resistance per square of the
normal region of the junction and the control channel is 0.4
$\Omega$, which yields, using free electron theory, an elastic
mean free path of 46 nm with diffusion constant D=0.02 $m^2/s$.
The Thouless energy, estimated from the junction dimensions, is
identical for both junctions and given by $E_{th}$=12$\mu eV$.

The geometry of the controllable Josephson junction used in the
controllable SQUID differs from the conventional cross shape. The
disadvantage is that the length L of the control channel is much
larger than in the case of a cross shaped device, resulting in a
diffusion time $\tau_D=\sqrt{L^2/D}\approx 2ns$. As a consequence,
a material with a long electron-electron relaxation time is needed
to be able to maintain a non-thermal energy distribution in the
control channel. For this reason silver is used as the normal
metal\cite{Pothier2,cucom}.

We now describe the sample fabrication, again  referring to Fig.
\ref{device}. The samples have been realized on a thermally
oxidized Si wafer that is covered with a 150 nm layer of sputter
deposited $Al_20_3$ to improve the adhesion of Ag. In the first
step the Nb ring is deposited using standard e-beam lithography on
a double layer of PMMA, DC sputtering and subsequent lift-off. The
critical temperature of the sputtered film is 8.1 K. Subsequently
the silver normal region, control channel and the reservoirs are
deposited in one single step using shadow evaporation. This is
needed because the adhesion of Ag is so poor that it is not
possible to bake this film to be able to do another lithography
step. We use a double layer of PMMA-MA and PMMA with e-beam
lithography and wet etching to create a PMMA suspended mask. The
deposition is done in an UHV deposition system with a background
pressure of $5 *10^{-10}$ mBar, the pressure in the system during
the evaporation steps is $\leq 5 *10^{-8}$ mBar. Prior to
deposition we use Argon etching ($P_{Ar}=1*10^{-4}$ mBar, 500 V)
for 3.5 minutes to clean the Nb surface. After that we deposit 10
nm of Ti adhesion layer under a large angle (47$^o$), with the
result that the Ti layer is only deposited on the substrate at the
position of the reservoirs, whereas it will be deposited on the
sides of the resist at the position of the thin openings defining
the control channel and the normal region of the junction.
Subsequently we deposit 50 nm Ag perpendicular to the substrate,
thus creating the control channel and the normal region of the
junction. As a last step 700 nm of Ag is deposited again at 47$^o$
to form the reservoirs with an effective thickness of 475 nm. To
measure the quality of the Nb-Ag interface we have made, in the
same run, a cross of a 200 nm wide Nb and Ag wire. The resistance
of the 200x200 nm interface has been determined to be 0.1 $\Omega$
which is smaller than the square resistance of the Silver (0.4
$\Omega$), indicating that the interface is clean. The SQUID shown
in Fig.\ref{device} has a normal state resistance of 0.55 $\Omega$
and, at 1.4 K, an equilibrium supercurrent ($\phi =0$, $V_{c,1}=
V_{c,2}=0$ mV) of 10 $\mu A$. The theoretical prediction of the
$\frac{I_cR_n}{E_{Th}}$ is 0.5, \cite{Dubos}, which corresponds
well with the measured value of $\frac{5.5}{12}=0.46$.

\begin{figure}
\centerline{\psfig{figure=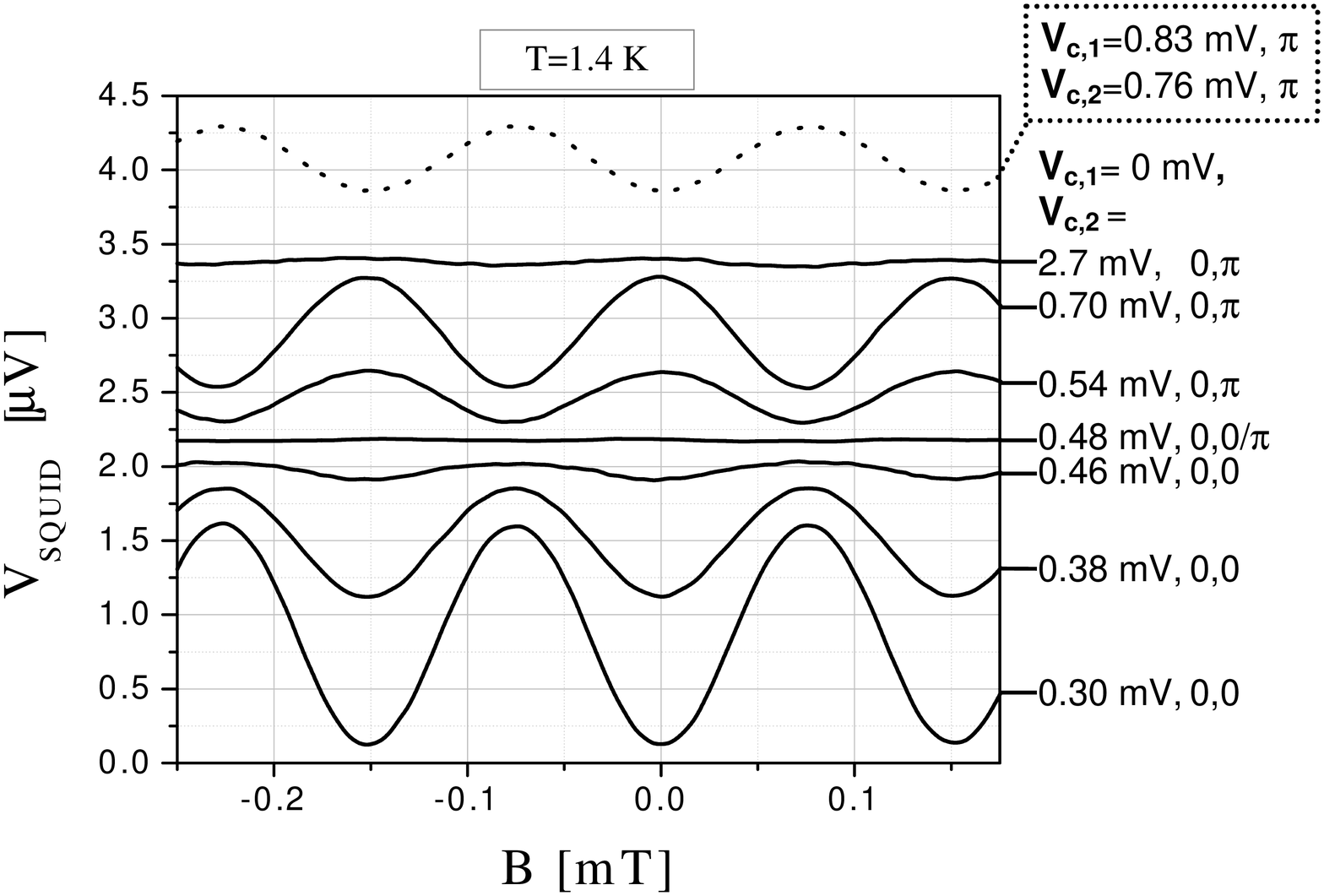,width=9 cm,clip=}}
\caption{The voltage over the controllable SQUID as a function of
the external magnetic field B for different values of $V_{c,1}$
and $V_{c,2}$ (curves offset for clarity).} \label{data1}
\end{figure}

In the experiment, we bias the SQUID with a low frequency AC bias
current ($I_{Bias}$,$~$80Hz) with an amplitude a little larger
than the critical current of the SQUID (see Fig \ref{device}) and
measure the voltage over the SQUID, $V_{SQUID}$, as a function of
the applied magnetic field B using a lock-in amplifier. This
lock-in technique strongly reduces the noise compared to a DC
biased measurement. Simultaneously we send a DC current through
the top and/or bottom control channel and measure the resulting
control voltage $V_{c,1}$ and/or $V_{c,2}$. Measurements are
performed at 100 mK, 1.4 K and at 4.2 K. A typical result, taken
at 1.4 K using the device shown in Fig \ref{device}, is shown in
Fig \ref{data1}. The solid lines represent the $V_{SQUID}$-B
oscillations for increasing values of $V_{c,2}$ ($V_{c,1}=0$)
using $I_{Bias}=4 \mu A$. At first the amplitude of the
oscillations decreases with increasing $V_{c,2}$ and reaches zero
at $V_{c,2}$=0.48 mV, indicating that the critical current of the
bottom junction is equal to 0. At higher values of $V_{c,2}$ the
$V_{SQUID}$-B oscillations re-appear, with a shift $\pi$ in phase
with respect to the oscillations at lower values of $V_{c,2}$. The
bottom junction and hence the SQUID, are now in the $\pi$ state.
At zero field we now measure a voltage maximum in stead of a
minimum, indicating that a circulating current is now flowing
around the SQUID loop. If the bottom junction is now kept in the
$\pi$-state ($V_{c,2}$=0.76 mV), and $V_{c,1}$ is increased to
0.83 mV, the top junction switches to the $\pi$-state as well.
This corresponds to an addition of 2 times $\pi$ to the phase of
the SQUID loop. In this case the original phase of the V-B
oscillations is regained, as shown by the dashed line in the
figure. Similar measurements at 100 mK in a dilution refrigerator
have shown similar results, with however larger amplitudes of the
V-B oscillations due to the temperature dependence of the critical
current of the Josephson junctions.

\begin{figure}
\centerline{\psfig{figure=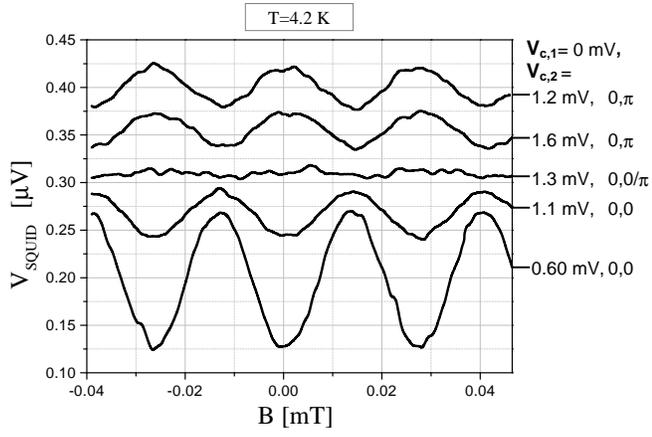,width=9 cm,clip=}} \caption{The
voltage over the controllable SQUID as a function of the external
magnetic field at 4.2K for different values of $V_{c,2}$ with
$V_{c,1}=0$ mV (curves offset for clarity).} \label{42}
\end{figure}

The question now arises whether the transition to a $\pi$-state
would be possible at 4.2 K. To be able to observe the effect at
these higher temperatures we have made another set of samples,
that differ only in the fact that the Josephson junctions are
shorter (length of the normal region: 870 nm, width 500 nm,
separation of the Nb electrodes: 260 nm, Rn=0.29 $\Omega$,
$E_{th}$=19 $\mu eV$) and that the surface area of the SQUID is
70.5 $\mu m^2$. We performed a measurement of the V-B oscillations
as a function of $V_{c,2}$ ($V_{c,1}$=0) at 4.2 K, with an AC
current bias of 1.5 $\mu$A. The results are shown in Fig.
\ref{42}. It is clear from the figure that despite a reduction in
the signal amplitude, due to the lower critical current and the
lower normal state resistance, the transition to the $\pi$ state
is observed at $V_{c,2}>1.3$ mV. This is a much higher value than
in the previous experiment, caused by the elevated temperature and
the larger Thouless energy. The observation of the $\pi$-state at
this temperature is somewhat surprising, for the transition to a
$\pi$-junction has so far only be observed at T$<$100 mK.

In summary, we have shown that it is possible to fabricate a
controllable $\pi$- SQUID, based on Nb-Ag, which operates in the
temperature range of 0.1 -4.2 K. The critical current of each
junction can be controlled by means of the application of a
control voltage $V_c$ over additional contacts attached to the
normal region of the specific junction. Moreover, the role of the
magnetic field, to apply $(n+\frac{1}{2})\phi_0)$ and thereby to
induce a circulating current in the SQUID, can be played by $V_c$,
which induces a screening current at integer external flux if its
value is large enough to cause the junction enter a $\pi$-state.

\begin{references}
\bibitem{cucom} We have made a SNS junction identical to the ones
used in the controllable SQUID using 50 nm of Cu as the normal
region and control channel. Only a monotonic decrease of the
supercurrent has been observed in this sample as a function of
$V_c$ (T=1.4K).
\end{references}
\end{document}